\documentclass[12pt]{article}
\usepackage{amsfonts,latexsym,eucal,color,graphicx,epsfig,amssymb,amsmath,cite}

\newcommand{\jj}{{\boldmath $\mathit{J}$}}
\newcommand{\be}{\begin{eqnarray}}
\newcommand{\ee}{\end{eqnarray}}

\begin{document}

\begin{titlepage}

\begin{flushright}
{\tt{arXiv:0804.1723 [hep-th]}}\\
\end{flushright}
\vspace{2cm}

\begin{center}
{\huge Curvature driven diffusion, Rayleigh-Plateau, and Gregory-Laflamme }
\end{center}
\vspace{1cm}

\begin{center}
\large{ Umpei Miyamoto }\\

\vspace{.5cm}
{\small {\textit{
Racah Institute of Physics, Hebrew University, Givat Ram, Jerusalem 91904, Israel 
}}
}
\\
\vspace*{0.5cm}

{\small
{\tt{
umpei@phys.huji.ac.il
}}
}\\
\end{center}
\vspace{1.0cm}

\begin{abstract}
It can be expected that the respective endpoints of the Gregory-Laflamme black brane instability and the Rayleigh-Plateau membrane instability are related because the bifurcation diagrams of the black hole-black string system and the liquid drop-liquid bridge system display many similarities. In this paper, we investigate the non-linear dynamics of the Rayleigh-Plateau instability in a range of dimensions, including the critical dimension at which the phase structure changes. We show that near the critical dimension and above, depending on a parameter in initial conditions an unstable cylinder will either pinch off or converge to an equilibrium state. The equilibrium state is apparently non-uniform but has a constant mean curvature everywhere. The results suggest that in the gravity side, near the critical dimension and above, the final state of an unstable black string (which is not too long) is a non-uniform black string. The equation of motion adopted to describe the dynamics is the surface diffusion equation, which was originally proposed to describe a grooving process of heated metal surfaces. An interesting correspondence between the diffusion dynamics and black hole (thermo)dynamics is discussed.
\end{abstract}

\end{titlepage}

%----------------------------------------------------------------------%
%----------------------------------------------------------------------%
\section{Introduction}
%----------------------------------------------------------------------%
%----------------------------------------------------------------------%

A variety of black hole solutions exist in higher dimensional spacetimes even with the asymptotic flatness condition~\cite{ER-review}. Furthermore, the compactification of an extra dimension has been known to make the phase structure richer~\cite{KK-review1,KK-review2}. For the spacetimes in which one spatial dimension is compactified on a circle, three phases of black objects are known, i.e., a localized black hole, a uniform black string, and a non-uniform black string~\cite{Harmark,Gubser,KW}. Although the phase diagrams, which are now available in 5 and 6 dimensions in the best form~\cite{KW}, suggest that the transition between them would occur, our knowledge about stability/dynamics of each phase is quite restricted. What we know is that the uniform black strings are unstable perturbatively if the linear dimension is too large as shown by Gregory and Laflamme~\cite{GL}. One might expect the endpoint of the instability to be a localized black hole since it has a larger entropy than the original uniform black string. Horowitz and Maeda, however, argued that the such a topology changing transition cannot occur due to the ``no-tear'' nature of the horizon and that the endpoint would be an non-uniform black string~\cite{HM1} (see also \cite{HM2}). Then, a general relativistic simulation was followed~\cite{chop,garf} and some suggestions were obtained, but no one has seen the unstable black string to pinch off nor any symptom of the convergence to an equilibrium state.

One of the most interesting aspects of the above black hole-black string system is that the phase structure seems to change drastically at a threshold spacetime dimension, $d=d_\ast:=14$~\cite{Sorkin}. In $d<d_\ast$, the uniform black string is thought to transit suddenly to a localized black hole (or probably also to a non-uniform black string, depending on dimension) as the compact dimension is adiabatically stretched with the mass kept fixed~\cite{Gubser,Kol2002}. While in $d\geq d_\ast$, the uniform black string is thought to transit smoothly to an infinitesimally deformed non-uniform black string, which would be stable unlike that in $d<d_\ast$. Thus, it is quite interesting to understand phase structures above/around the critical dimensions and to know whether or not the stability of non-uniform black string indeed changes around the critical dimension. We note that the number of a critical dimension depends on the type of ensemble~\cite{KM}, momentum~\cite{hov}, and charge~\cite{MK}. For example, the critical dimension in a canonical ensemble is $\tilde{d}_\ast = 13$.

The membrane paradigm of black holes~\cite{mempara} often provides useful tools to investigate how black holes actually behave and a new perspective for the black hole physics~\cite{Kovtun:2003wp}. In the context of the black hole-black string system, a lot of similarities between the Gregory-Laflamme instability and the Rayleigh-Plateau instability, which is a universal instability of extended fluids/membranes~\cite{Plat,Rayl}, were pointed out perturbatively~\cite{CD,CG,CDG}. Then, the previous paper~\cite{MM} by the present author and a collaborator revealed that the similarities of the phase structures between the black hole-black string system and a liquid drop-liquid bridge system persist up to non-linear regimes. In particular, it was shown that a critical dimension similar to that in the gravity side~\cite{Sorkin} exists in the fluid side.
That is, the phase diagram implies that a Uniform Bridge (UB) suddenly transits to either a Non-Uniform Bridge (NUB, known as the \textit{Delaunay unduloid}~\cite{unduloid}) or a Spherical Drop (SD) below a critical space dimension, $D < D_\ast:=12 $, as the compact dimension is adiabatically stretched with the volume kept fixed. While in $D \geq D_\ast$, the phase diagram implies that the UB smoothly transits to an infinitesimally deformed NUB just like the transition of the uniform black string to non-uniform black string in $d\geq d_\ast$.

In the previous paper~\cite{MM}, the equilibria of a liquid were obtained by the variational calculus which minimizes/extremalizes the surface area of the liquid while keeping the volume kept fixed (i.e., \textit{Plateau's problem} or the \textit{capillary minimization problem}). Thus, a liquid surface obtained is a minimal surface, which has a \textit{constant mean curvature} everywhere. The next step would be to establish the (in)stability of each phase. Although one can investigate the local stability by the second variation of an area functional, it may be more interesting to see directly the global (in)stability adopting a suitable dynamical equation of motion. What kind of dynamics is appropriate?\ A necessary condition for a candidate is to have the equilibria represented by the constant mean curvature surfaces. In addition, it should be required that the motion decreases the surface area while keeping the volume fixed throughout time evolutions. Is there such a convenient dynamical equation?\ Yes, the \textit{surface diffusion equation}, which was originally proposed to describe a grooving process of heated polycrystal surfaces in metallurgy~\cite{Mullins}, does have all properties raised above. Within the framework of the surface diffusion, it is known that in the usual 3 dimensional (3D) space the SD is stable, the NUB is unstable, and the UB is unstable (the Rayleigh-Plateau instability) if its linear dimension is longer that its circumference. It is also known that the unstable UB dynamically transits to the SD via a topology changing transition in a self-similar manner~\cite{CFM,BBW} (see \cite{BBW} for a comprehensive study of axisymmetric dynamics).

In this paper, we investigate the linear and non-linear properties of the Rayleigh-Plateau instability and especially their endpoints in the surface diffusion dynamics. We will show (i) that the Rayleigh-Plateau instability has a dimensional dependence similar to that of the Gregory-Laflamme instability; (ii) an interesting aspect of the Rayleigh-Plateau instability appearing in higher-order perturbations; (iii) that an unstable UB pinches off in lower dimensional spaces (e.g., in 4D), while the unstable long (but not too long) UB converges to the NUB in higher dimensional spaces (e.g., in 12D); and point out (iv) a remarkable correspondence between the surface diffusion dynamics and the black hole thermodynamics.

%----------------------------------------------------------------------%
%----------------------------------------------------------------------%
\section{Axisymmetric Surface Diffusion}
\label{sec:diff}
%----------------------------------------------------------------------%
%----------------------------------------------------------------------%

%----------------------------------------------------------------------%
%----------------------------------------------------------------------%
\subsection{ Basic Equation }
\label{sec:axi}
%----------------------------------------------------------------------%
%----------------------------------------------------------------------%

The surface diffusion is a mass transport phenomenon on material surfaces driven by the gradient of a curvature~\cite{Mullins}. A superficial flux of particle numbers is given via the Nernst-Einstein relation by
\be
	\mbox{\jj}
	=
	- A \nabla_{\!\! s} \kappa\ ,
\ee
where $A$ is a constant, $ \nabla_{\!\! s} $ the surface Laplacian, and $ \kappa $ the mean curvature of the surface. Local volume conservation results in the motion of the surface:
\be
	\rho u + \nabla_{\!\! s} \cdot \mbox{\jj} = 0\ ,
\ee
where $ u $ is the normal velocity of the surface and $ \rho $ is a constant density of atom numbers. Thus, the motion of surface obeys the surface diffusion equation,\footnote{In the original theory of the thermal grooving~\cite{Mullins}, the constant $B$ is given by $B = D_s \gamma v^2 \nu / k_B T$, where $D_s$ is the coefficient of surface diffusion, $\gamma$ the surface-free energy per unit area, $v=\rho^{-1}$ the molecular volume, $\nu$ the number of atoms per unit area, and $T$ the temperature.}
\be
	u
	=
	\rho^{-1} A \nabla_{\!\! s}^2 \kappa
	=:
	B \Delta_{s} \kappa\ .
\label{eq:pre-eom}
\ee

\begin{figure}[t]
	\begin{center}
		\includegraphics[width=8cm]{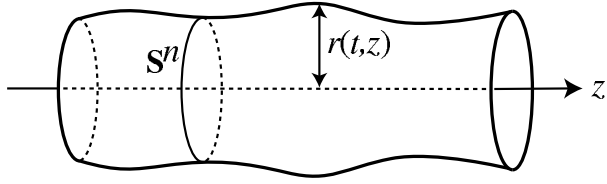}
		\caption{
{\footnotesize An axisymmetric hypersurface in $(n+2)$-dimensional space }
				}
		\label{fg:rz}
	\end{center}
\end{figure}

From now on we concentrate on the motion of axisymmetric $(n+1)$-dimensional ``metal surfaces'' in $ D:=(n+2) $-dimensional space $ (n\geq 1) $. We denote the instantaneous radius of $ \mathbf{S}^n $ at $z$ by $ r = r(t,z) $, where $t$ is the time and $z$ is the axis of symmetry (see Fig.~\ref{fg:rz}). In this coordinates, the mean curvature of the surface is given by
\be
	\kappa
	=
	\frac{ n }{ r \sqrt{ 1+r^{\prime 2} } }
	-
	\frac{ r^{\prime\prime} }{ ( 1+r^{\prime 2} )^{3/2} }\ ,
\label{eq:kappa}
\ee
where $X^\prime = \partial_z X$. See Appendix for the calculation detail. The first and second terms in Eq.~(\ref{eq:kappa}) have the meanings of the azimuthal and axial principal curvatures, respectively. The normal velocity and the surface Laplacian are given by
\be
	u = \frac{ \partial_t r }{ \sqrt{ 1+r^{\prime 2} } }\ ,
\;\;\;\;\;\;
	\Delta_{s}
	=
	\frac{ 1 }{ r^n \sqrt{ 1+r^{\prime 2} } } \partial_z
	\left(
		\frac{ r^n }{ \sqrt{ 1+r^{\prime 2} } } \partial_z
	\right)\ .
\label{eq:nabla}
\ee
The combination of Eqs.~(\ref{eq:pre-eom}), (\ref{eq:kappa}), and (\ref{eq:nabla}) gives the basic equation governing the dynamics of axisymmetric surface diffusions
\be
	\partial_t r
	=
	\frac{ B }{ r^n } \partial_z  
	\left(
		\frac{ r^n  }{ \sqrt{ 1+r^{\prime 2} } } \partial_z
			\left[
				\frac{ n }{ r \sqrt{ 1+r^{\prime 2} } }
				-
				\frac{ r^{\prime\prime} }{ ( 1+r^{\prime 2} )^{3/2} }
			\right]
	\right)\ ,
\label{eq:eom}
\ee
which is first order and fourth order with respect to $t$ and $z$, respectively.
It is important to notice that from Eq.~(\ref{eq:pre-eom}) the constant mean curvature surfaces ($\kappa=\mathrm{const.}$), e.g., the Uniform Bridge (UB), Non-Uniform Bridge (NUB), and Spherical Drop (SD) obtained in~\cite{MM}, are stationary solutions of the surface diffusion equation.

%----------------------------------------------------------------------%
%----------------------------------------------------------------------%
%----------------------------------------------------------------------%
%----------------------------------------------------------------------%

%----------------------------------------------------------------------%
%----------------------------------------------------------------------%
\subsection{ Volume Conservation and Area Decreasing }
\label{sec:vol}
%----------------------------------------------------------------------%
%----------------------------------------------------------------------%

Now, let us see two important aspects of the surface diffusion dynamics, i.e., the conservation of volume and the decreasing of surface area. When we consider an object extending in a region of $z \in P := [-L/2,L/2]$, the instantaneous volume and surface area of the body can be written as
\be
&&
V(t)
	=
	\Omega_n \int_P dz\;
	r^{n+1}(t,z)\ ,
\nonumber
\\
&&
	A(t)
	=
	(n+1) \Omega_n \int_P dz
	\sqrt{ 1+r^{ \prime 2 } }\; r^n(t,z)  \ .
\label{eq:avdot}
\ee
Here, $\Omega_n := \pi^{ (n+1)/2 }/\Gamma[(n+1)/2+1]$ is the volume of a unit $n$-sphere.
Taking a time derivative of these equations, one can show
\be
&&
	\dot{V}(t)
	=
	B (n+1) \Omega_{n}
	\left. 
	\frac{ r^n \kappa^\prime }{ \sqrt{ 1+r^{ \prime 2} } } \right|_{\partial P} \ ,
\nonumber
\\
&&
	\dot{A}(t)
	=
	-
	B (n+1) \Omega_n \int_P dz\;
	\frac{ r^n \kappa^{ \prime 2 } } { \sqrt{ 1 + r^{ \prime 2} } }
	+
	(n+1) \Omega_n
	\left.
		\frac{ r^n ( B \kappa \kappa^\prime +  r^\prime \partial_t r  )	}
			 { \sqrt{ 1+r^{ \prime 2} } }
	\right|_{\partial P} \ .
\label{eq:dot}
\ee
From equations in (\ref{eq:dot}), we see that the volume does not change ($\dot{V}=0$) and the surface area decreases ($\dot{A}\leq 0$) throughout a time evolution if the following periodic boundary conditions hold,
\be
	r^{(l)}(t,-L/2)
	=
	r^{(l)}(t,L/2)\ ,
\;\;\;\;\;
	\kappa^{(l)}(t,-L/2)
	=
	\kappa^{(l)}(t,L/2)\ ,
\;\;\;\;\;
	(l=0,1)\ .
\label{eq:bc}
\ee
Here, $ X^{(l)}:= \partial_z^l X $. The equality $\dot{A}=0$ holds when the hypersurface has a homogeneous mean curvature, $\kappa^\prime=0$.

%----------------------------------------------------------------------%
%----------------------------------------------------------------------%
\subsection{Rayleigh-Plateau Instability}
\label{sec:RP}
%----------------------------------------------------------------------%
%----------------------------------------------------------------------%

\begin{figure}[t]
	\begin{center}
		\includegraphics[width=8cm]{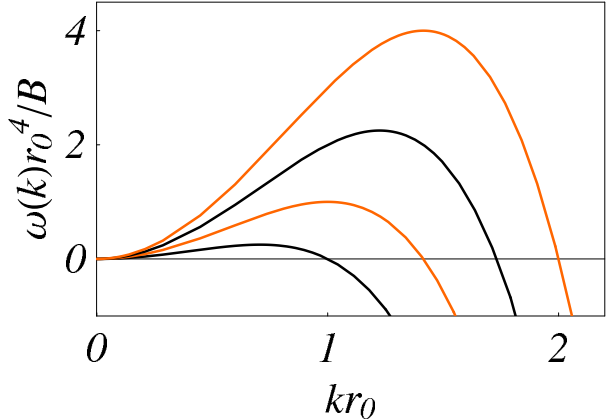}
		\caption{
\footnotesize{
Growth rate of the Rayleigh-Plateau instability from 3D (bottom curve) to 6D (top curve).
			 }
				}
		\label{fg:disp}
	\end{center}
\end{figure}

The linear perturbation around a UB sets a starting point of the investigation of dynamics. We expand $r(t,z)$ around the UB as $r(t,r) = r_0 + \varepsilon r_1(t,z)$ with a small expansion parameter $\varepsilon$ and a radius of UB $r_0$. Substituting this expansion into Eq.~(\ref{eq:eom}), we have the linear perturbation equation at $O(\varepsilon)$,
\be
	\partial_t r_1
	+
	B
	\left(
		\frac{n}{r_0^2} r_1^{ \prime\prime }
		+
		r_1^{ (4) }
	\right)
	= 0 \ .
\label{eq:first}
\ee
Now, we consider a superposition of $N$ sinusoidal waves as an initial condition given by
\be
r_1(0,z) = \sum_{i=1}^{N} a_i \cos (k_i z)\ ,
\label{eq:r1-ini}
\ee
where $a_i$ and $k_i$ are constants. Then, the perturbative equation~(\ref{eq:first}) is solved to give
\be
&&
	r_1 (t,z)= \sum_{i=1}^{N} a_i e^{ \omega(k_i) t }  \cos(k_i z)\ ,
\label{eq:r1}
\\
&&
	\omega(k)
	:=
	\frac{ B }{ r_0^4 } (k r_0)^2 \left[ n - ( k r_0)^2   \right] \ .
\label{eq:disp}
\ee
Equations~(\ref{eq:r1}) and (\ref{eq:disp}) tell us that the initial perturbation of wavenumber $k_i$ grows exponentially if $ 0< k_i < k_{\mathrm{cr}}:= \sqrt{n}/r_0$ for a given $r_0$, which implies the Rayleigh-Plateau instability is a long wavelength instability just like the Gregory-Laflamme instability. Furthermore, the dimensional dependence of this critical wavenumber is similar to that of the Gregory-Laflamme instability, and they indeed coincide each other in the large dimension limit, $n\to +\infty$~\cite{KS,vadim} (see also \cite{CD}). We plot the dispersion relation~(\ref{eq:disp}) in Fig.~\ref{fg:disp}. The peak of growth rate, $ \omega_{\mathrm{max}}:=\omega(\sqrt{n/2}/r_0)=Bn^2/4r_0^2 $, also increases with dimension, which is similar to the Gregory-Laflamme instability again~\cite{GL}. One difference between two instabilities seems to exist near the infrared region $k r_0 \ll 1$, where the growth rate of the Rayleigh-Plateau instability is suppressed faster than the Gregory-Laflamme instability. This behavior would come from the ``short distance nature'' of the curvature driven surface diffusion, relating to the fourth derivative term in Eq.~(\ref{eq:first}). Note that the dimensional dependence of the critical wavenumber, $k_{\mathrm{cr}}\propto \sqrt{n}$, comes not from the Laplacian in Eq.~(\ref{eq:nabla}) but from the azimuthal principal curvature in Eq.~(\ref{eq:kappa}). Therefore, we can say that the dimensional dependence comes from that of the background curvature of $\mathbf{S}^n$ rather than that of the ``interaction'' represented by the higher-order derivatives.

One can generalize the above linear perturbation of UB to higher-order ones. In Appendix, we present the result of second-order perturbation. Here, we just mention an interesting aspect of Rayleigh-Plateau instability peculiar to the non-linear regime. If we set $k_i > \sqrt{n}/r_0$, ($1\leq i \leq N$) in the initial condition (\ref{eq:r1-ini}), the linear perturbation analysis tells us that the initial perturbations will be exponentially suppressed. Even in this case, however, a certain class of modes in the second-order perturbation \textit{grows} if there exists a couple of sinusoidal waves satisfying $| k_i - k_j | < \sqrt{n}/r_0 $, ($ i \neq j $) in the initial perturbation~\cite{CFM}.

%----------------------------------------------------------------------%
%----------------------------------------------------------------------%
%----------------------------------------------------------------------%
%----------------------------------------------------------------------%

%----------------------------------------------------------------------%
%----------------------------------------------------------------------%
\section{Non-Linear Dynamics}
\label{sec:dyn}
%----------------------------------------------------------------------%
%----------------------------------------------------------------------%

%----------------------------------------------------------------------%
%----------------------------------------------------------------------%
\subsection{ Boundary and Initial Conditions }
\label{sec:bc}
%----------------------------------------------------------------------%
%----------------------------------------------------------------------%

Now, we are ready for the numerical investigation to know the endpoint of the Rayleigh-Plateau instability. The partial differential equation (\ref{eq:eom}) is solved for various $n$. We solve the equation in the region of $z\in P = [-L/2,L/2]$ with the periodic boundary condition (\ref{eq:bc}), which is equivalent to
\be
	r^{(l)}(t,-L/2)
	=
	r^{(l)}(t,L/2)\ ,
\;\;\;\;\;
	(l=0,1,2,3)\ .
\label{eq:bc2}
\ee
With boundary condition (\ref{eq:bc2}), one can model the motion of an infinitely long periodic body. For simplicity, a UB given a single sinusoidal perturbation is chosen as the initial condition:
\be
	r(0,z)
	=
	r_0 \left[ 1 + \varepsilon \cos \left( p z \right) \right]\ ,
\;\;\;\;
	p = 2\pi/L \ ,
\;\;\;\;
	L
	=
	(1+\delta) L_{\mathrm{cr}}\ ,
\label{eq:IC}
\ee
where $ L_{ \mathrm{cr} } := 2\pi / k_{ \mathrm{cr} } $ is the critical wavelength, $\varepsilon$ and $\delta$ are small numbers ($\delta>0$). Thus, we have a couple of free parameters $(\varepsilon,\delta)$ in our minimal-setting simulation. In the following numerical calculation, we will set these parameters to
\be
	\varepsilon = -0.10\ ,
	\;\;\;\;
	\delta = 0.12\ .
\label{eq:paras}
\ee
The positivity of $\delta$ implies that the initial perturbation has a wavenumber smaller than the critical wavenumber ($p < k_{ \mathrm{cr}}$ or equivalently $L>L_{\mathrm{cr}}$) and will grow at least during an early time of the evolution. A typical dynamical time scale can be defined using the growth rate of linear perturbation (\ref{eq:disp}) as
\be
	t_{\mathrm{dyn}}
	:=
	1/\omega(p)
	\ .
\ee
For example, one can normalize length and time by $r_0$ and $t_{\mathrm{dyn}}$, respectively.

%----------------------------------------------------------------------%
%----------------------------------------------------------------------%
\begin{figure}[b]
	\begin{center}
		\setlength{\tabcolsep}{ 15 pt }
		\begin{tabular}{ ll }
			\includegraphics[width=6cm]{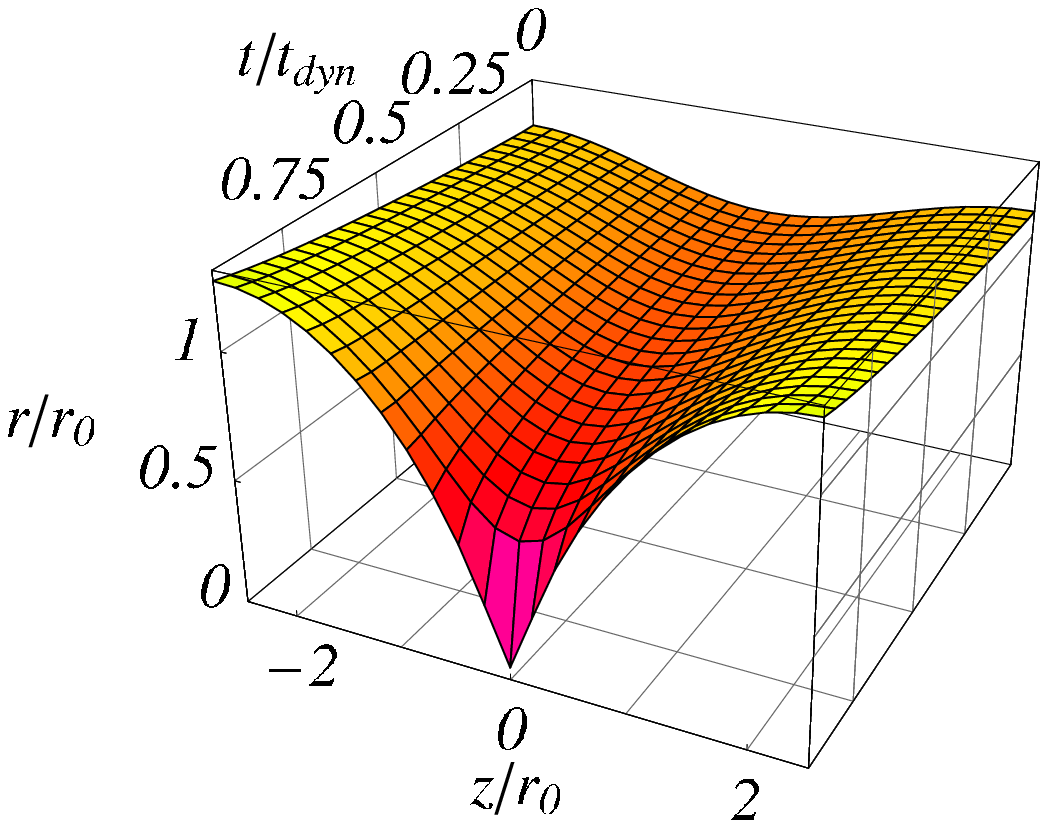} &
			\includegraphics[width=6cm]{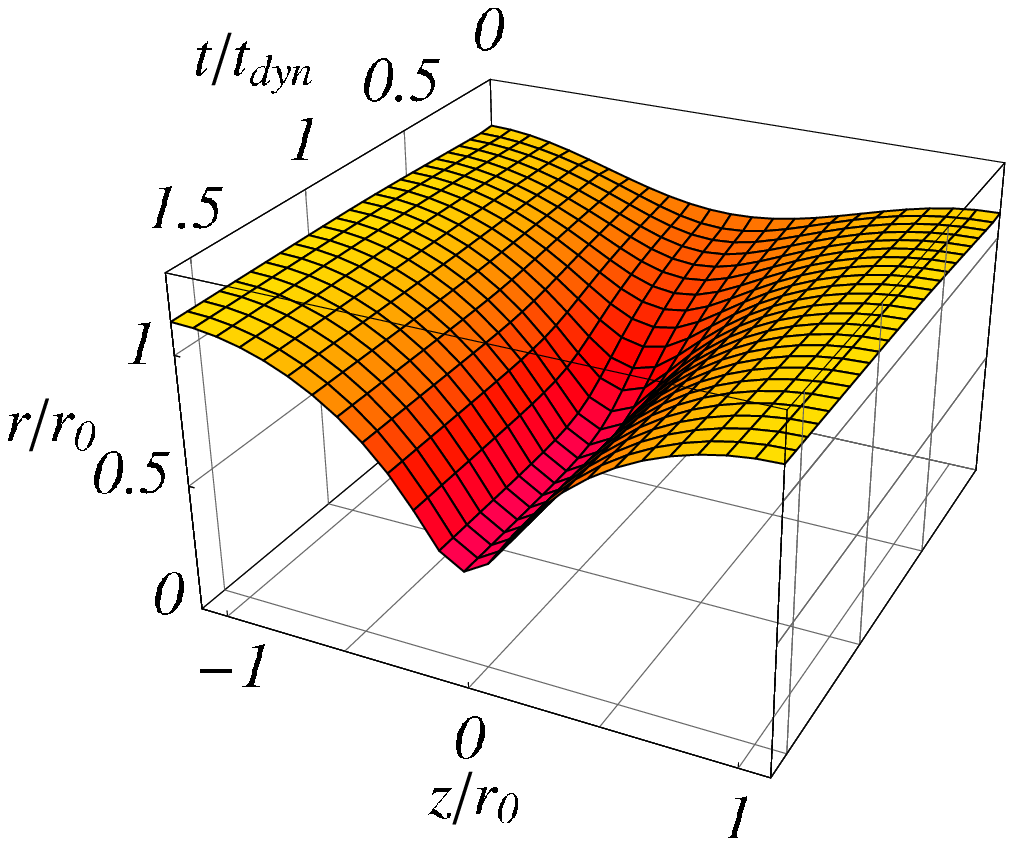} \\
		\end{tabular}
	\caption{
\footnotesize{
Time evolution of $r=r(t,z)$ in 4D (left) and 12D (right).
}
			} \label{fg:evo}
	\end{center}
\end{figure}
%----------------------------------------------------------------------%
%----------------------------------------------------------------------%

%----------------------------------------------------------------------%
%----------------------------------------------------------------------%
\begin{figure}[t]
	\begin{center}
		\setlength{\tabcolsep}{ 3 pt }
		\begin{tabular}{ llll }
			$ t/t_{\mathrm{dyn}}  = 0.0 $ &
			$ t/t_{\mathrm{dyn}}  = 0.85 $ &
			$ t/t_{\mathrm{dyn}}  = 0.91 $ &
			$ t/t_{\mathrm{dyn}}  = 0.92 $ \\
			\includegraphics[width=4cm]{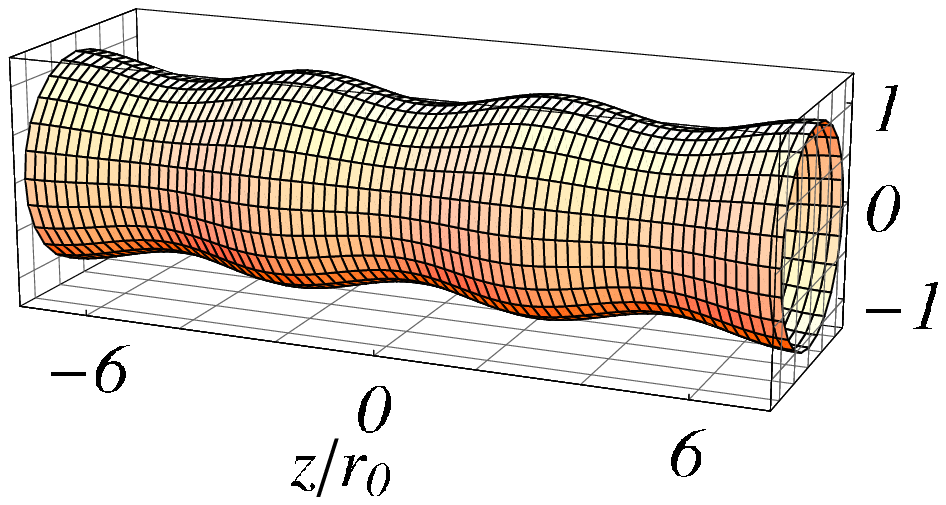} &
			\includegraphics[width=4cm]{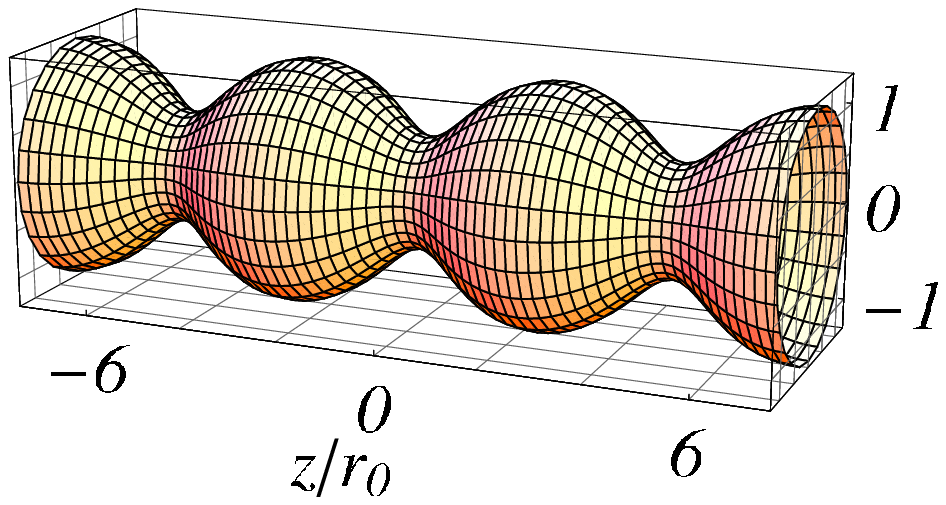} &
			\includegraphics[width=4cm]{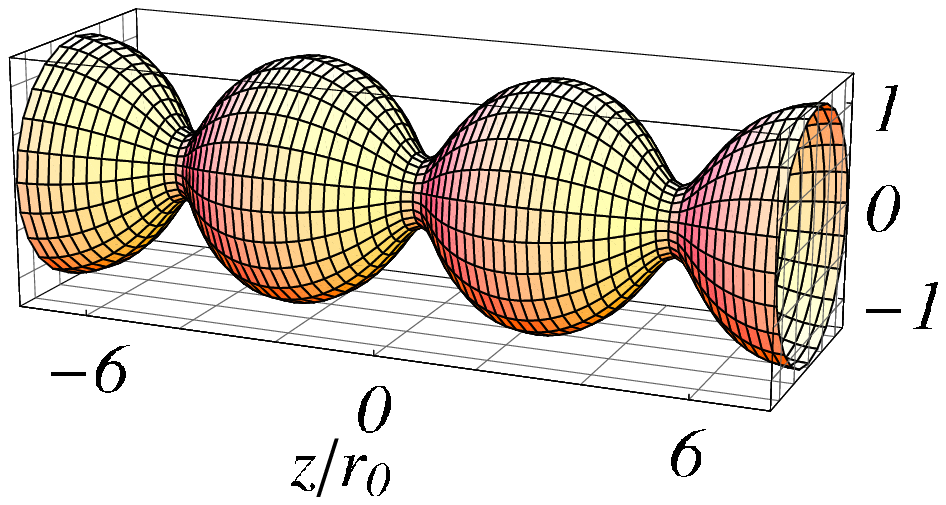} &
			\includegraphics[width=4cm]{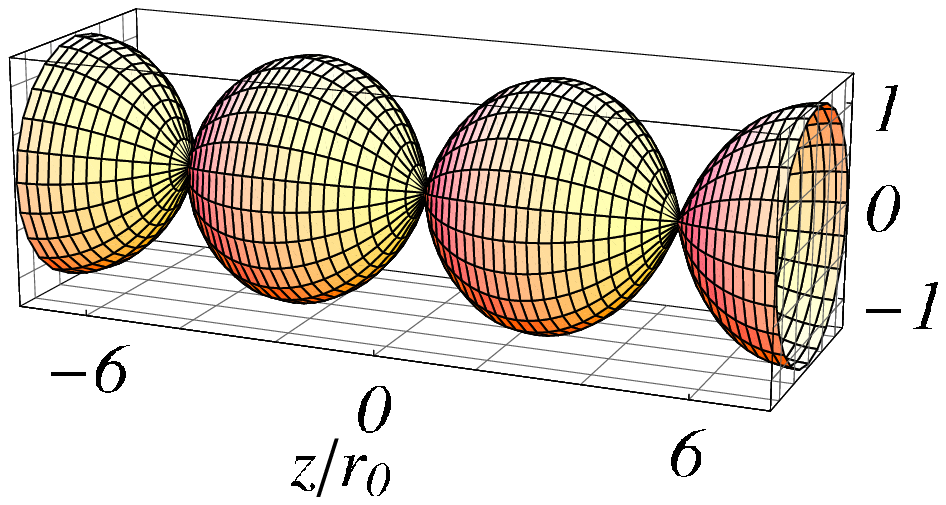} \\
			$ t/t_{\mathrm{dyn}}  = 0.0 $ &
			$ t/t_{\mathrm{dyn}}  = 0.42 $ &
			$ t/t_{\mathrm{dyn}}  = 0.67 $ &
			$ t/t_{\mathrm{dyn}}  = 1.18 $ \\
			\includegraphics[width=4cm]{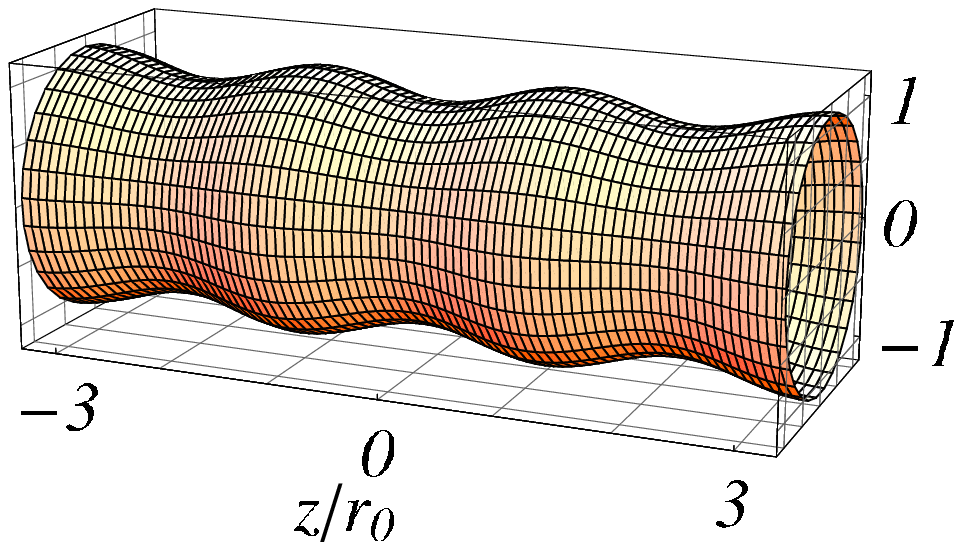} &
			\includegraphics[width=4cm]{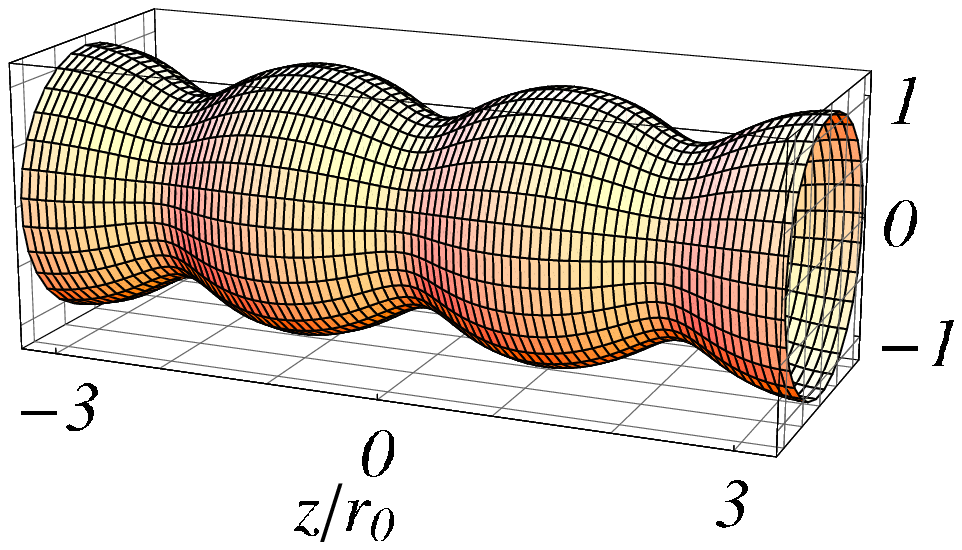} &
			\includegraphics[width=4cm]{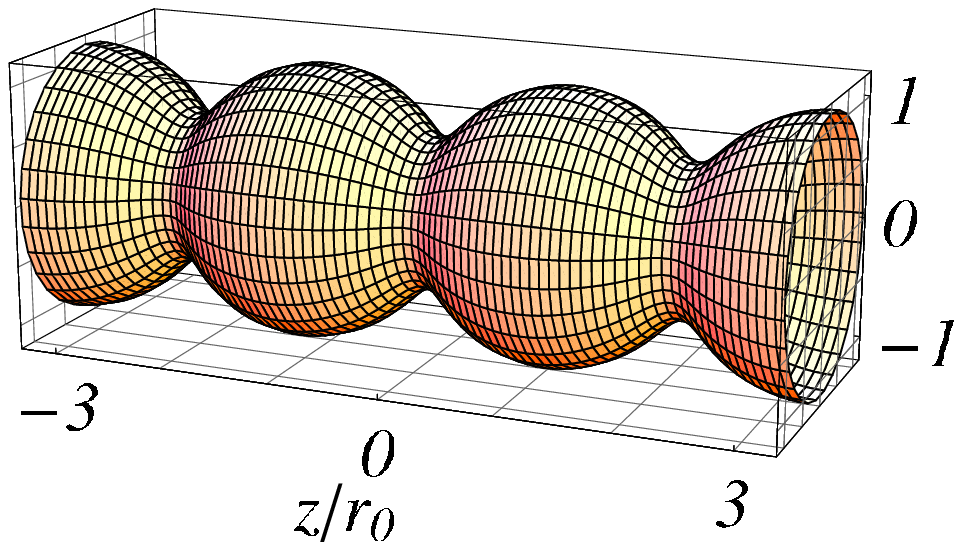} &
			\includegraphics[width=4cm]{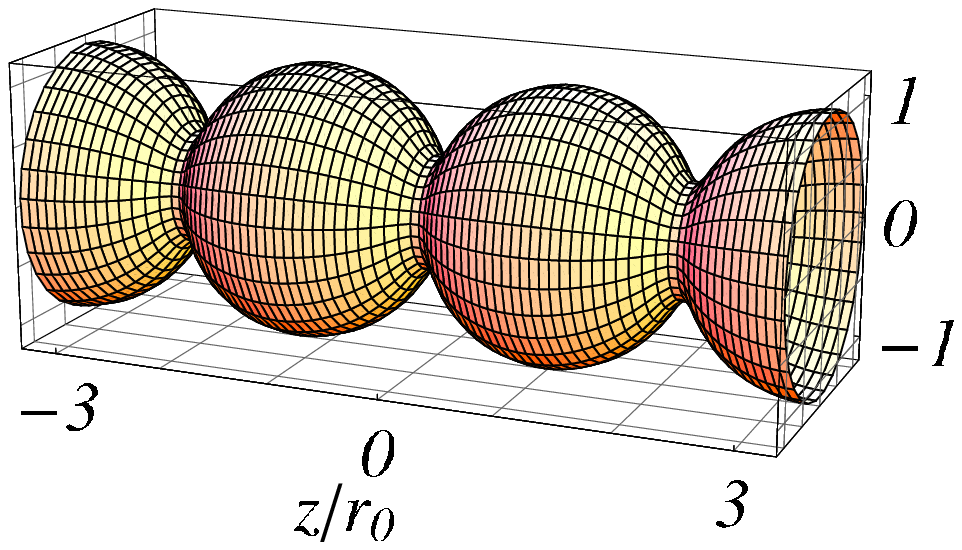} \\
		\end{tabular}
	\caption{
\footnotesize{
Snapshots at selected moments of time in 4D (top) and 12D (bottom).
The evolution results in pinching off in 4D and convergence to a non-uniform equilibrium state in 12D. The final equilibrium (12D) is apparently highly non-uniform but has a constant mean curvature everywhere (see Fig.~\ref{fg:kappa}).
}
			} \label{fg:conf}
	\end{center}
\end{figure}
%----------------------------------------------------------------------%
%----------------------------------------------------------------------%

%----------------------------------------------------------------------%
%----------------------------------------------------------------------%
\subsection{ Fate of Instability in 4D and 12D }
\label{sec:fate}
%----------------------------------------------------------------------%
%----------------------------------------------------------------------%

We will show typical behaviors taking two examples, $ D = n+2 = 4 $ $(< D_\ast)$ and $D=n+2=12$ $(=D_\ast)$. The time evolution of $r=r(t,z)$ obtained numerically is shown in Fig.~\ref{fg:evo} for these cases.

In 4D, the given initial perturbation continues to grow and the numerical solution suggests the breakup of the body to become a pair of separated drops near $t/t_{\mathrm{dyn}} \simeq 0.92 $, just like the behavior observed in the 3D axisymmetric surface diffusion~\cite{CFM,BBW}. Since the breakup, represented by $r=0$, is a singularity in our coordinates, we have to change variables before the breakup. Although the continuation of calculation through the breakup is possible with some prescriptions~\cite{CFM,BBW}, we simply stopped the calculation since to know the detail of the breakup is not the purpose of this paper. Nevertheless, we \textit{can} argue that the numerical result implies the pinch off from the behavior of mean curvature, as we will see soon. In 12D, the given initial perturbation grows during an early time, which is consistent with the linear- and second-order perturbations. However, the numerical solution suggests that the surface begins to converge to a certain non-uniform configuration due to non-linear effects. Several snapshots of the hypersurface at selected moments of time are shown in Fig.~\ref{fg:conf}.

\begin{figure}[t]
	\begin{center}
		\setlength{\tabcolsep}{ 15 pt }
		\begin{tabular}{ cc }
		\includegraphics[width=6cm]{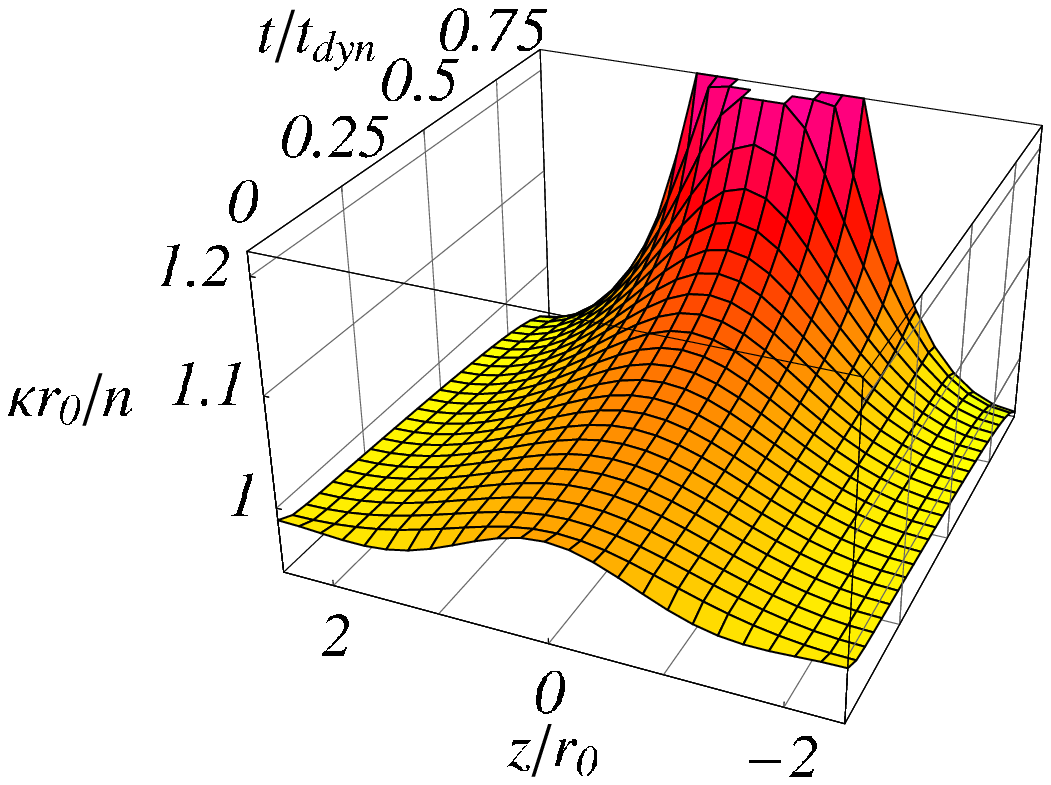} &
		\includegraphics[width=6cm]{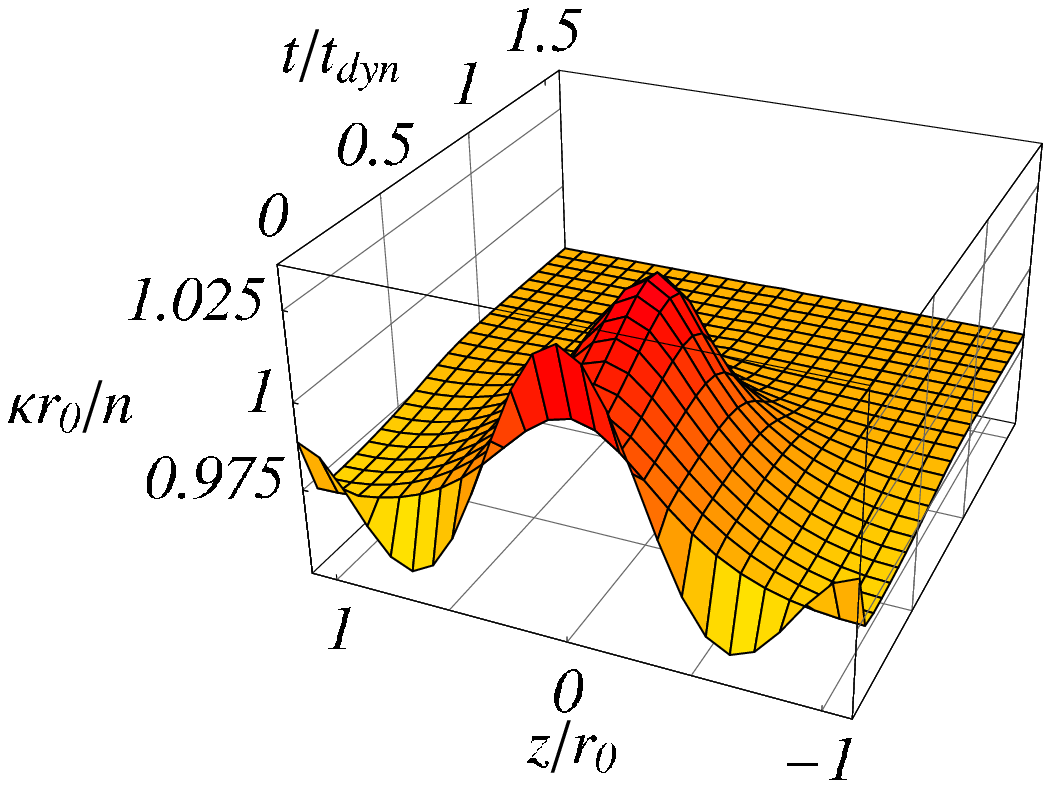} 
		\end{tabular}
	\caption{
\footnotesize{
Time evolution of the mean curvature $\kappa=\kappa(t,z)$ in 4D (left) and 12D (right). Note that the viewpoint is different from that in Fig.~\ref{fg:evo}.
}
		} \label{fg:kappa}
	\end{center}
\end{figure}

Now, we see the evolution of the mean curvature $\kappa$. The time evolution of $\kappa$ is shown in Fig.~\ref{fg:kappa} for both dimensions. Note that in both cases the initial sinusoidal perturbation breaks the constancy of the mean curvature of the background UB, $ \kappa_{ \mathrm{UB} } = n/r_0 $. In 4D, the mean curvature continues to grow with an increase of spatial gradient especially around the center, $z=0$. This means that the current $ |\mbox{\jj}| $ continues to grow and the surface deviates from an equilibrium state. This behavior supports our interpretation that the surface pinches off and also approaches the other equilibrium state, i.e., the SD which has a smaller area~\cite{MM}. While in 12D, the mean curvature becomes a global constant after a few oscillations localized near the center. This behavior implies the surface converges to the equilibrium state having a constant mean curvature, i.e., the NUB obtained in the previous paper~\cite{MM}.

%----------------------------------------------------------------------%
%----------------------------------------------------------------------%
%----------------------------------------------------------------------%
%----------------------------------------------------------------------%

%----------------------------------------------------------------------%
%----------------------------------------------------------------------%
\subsection{ Initial Condition Dependence and Fate in Other Dimensions }
\label{sec:fate}
%----------------------------------------------------------------------%
%----------------------------------------------------------------------%

To state criteria for a body to pinch off or converge to a NUB, it is convenient to introduce the following reduced volume
\be
	\mu := \frac{ V }{ L^{ n+2 } }\ ,
\label{eq:mu}
\ee
which is invariant under the overall scaling of the system, $L\to\beta L$ and $V\to \beta^{n+2}V$. Since the volume is conserved and the period $L$ is fixed in our simulations, $\mu$ remains a constant throughout the evolution. Each class of constant mean curvature surfaces such as that of UBs, NUBs, and SDs is parameterized by $\mu$~\cite{MM}. In particular, the stability analysis of the UB says that a UB parameterized by $\mu$ is unstable if $\mu<\mu_{\mathrm{cr}}=\Omega_n (\sqrt{n}/2\pi)^{n+1}$.

In the 4D case, the phase diagram of constant mean curvature surfaces~\cite{MM} tells us that the SD, which would be the endpoint of the dynamics, exists even for $\mu < \mu_{\mathrm{cr}}$ with an area smaller than that of the UB having the same $\mu$. This suggests the pinch off always happens whenever we take $\mu<\mu_{\mathrm{cr}}$ initially, i.e., take $ \delta > 0 $ in the initial condition~(\ref{eq:IC}). We confirmed this by taking several values of $\delta$ within the accuracy of our calculation. While in the 12D case, the phase diagram~\cite{MM} tells us that the NUB, which would be the endpoint of the dynamics, exists only in $ 0.28 \mu_{\mathrm{cr}} \lesssim \mu < \mu_{\mathrm{cr}}$ with an area smaller than that of the UB having the same $\mu$. This means that if we take a large value of $ \delta $ resulting in $ \mu < 0.28 \mu_{\mathrm{cr}}$, the perturbed UB cannot converge to the NUB but pinches off and approaches a SD, which exists for such a small $\mu$. We confirmed this (in the sense described in the previous subsection) by taking several values of $\delta$ greater than that in Eq.~(\ref{eq:paras}). The parameter $\varepsilon$ does not play any important role since this quantity just changes the dynamical time scale.

Here, we mention the fate of Rayleigh-Plateau instability in other dimensions within $3 \leq D \leq 15$. For the pair of $(\varepsilon,\delta)$ in Eq.~(\ref{eq:IC}), we observed the pinch off in $3 \leq D \leq 11$ and the convergence to the NUB in $ 12 \leq D \leq 15$, while the non-uniformness of the final equilibrium depends on dimension. We should note that if we take $ \delta $ too small ($0<\delta\ll 1$), the UB converges to the NUB even in 10D and 11D since there exists the NUB branch having an area smaller than that of the UB~\cite{MM}.

%----------------------------------------------------------------------%
%----------------------------------------------------------------------%
\section{Discussion}
\label{sec:disc}
%----------------------------------------------------------------------%
%----------------------------------------------------------------------%

We have seen that the non-uniform bridge indeed can serve as the global attractor of the surface diffusion dynamics associated with the Rayleigh-Plateau instability. Combining this result with the correspondence of the phase diagrams between the black hole-black string and liquid drop-liquid bridge systems~\cite{KW,MM}, it is likely that the fate of the unstable long (but not too long) black string is the non-uniform black string around the critical dimension and above. On the other hand, although we have seen that the Rayleigh-Plateau instability leads the unstable uniform bridge to pinching off in lower dimensions, the pinch off of an event horizon will be associated with a naked singularity and other problems peculiar to curved spacetimes. Therefore, the relation between the respective endpoints needs further considerations in such lower dimensional cases\footnote{If the evolution results in pinching off also in the black string dynamics, a certain class of multi-black hole configurations should be interpreted as the endpoint of the Gregory-Laflamme instability~\cite{Dias}.}.

The existence of critical dimensions seems to stem from purely geometrical or morphological effects not peculiar to the black strings and minimal surfaces. Therefore, the convergence to non-uniform configurations would be realized in other physical systems in which an attractive force governs the dynamics. It is interesting to find the convergence to non-uniform configurations, for instance, in the Rayleigh-Plateau instability in a hydrodynamical system, the Dyson-Chandrasekhar-Fermi instability as a Newtonian gravity counterpart of the Gregory-Laflamme instability~\cite{CG,CD}, and a system of gravitating shell/domain wall.

One who is familiar with the black hole thermodynamics~(e.g., \cite{novikov}) might wonder that the ``one-way nature'' of the surface area ($\dot{A}\leq 0$) brings to mind the second law of the black hole thermodynamics, i.e., the area of an event horizon increases ($\delta A_{\mathrm{EH}} \geq 0$) or the free energy of a black hole decreases ($\delta F \leq 0$). Furthermore, that the diffusion dynamics converges to an equilibrium state if and only if the mean curvature $\kappa$ becomes a global constant is quite similar to the zeroth law, which states the global constancy of the surface gravity on event horizons for stationary black holes\footnote{In the theory of surface diffusion, the mean curvature $\kappa$ originates from the chemical potential on metal surfaces rather than the temperature~\cite{Mullins}, while the surface gravity of a black hole is proportional to the black hole temperature.}. Thus, the surface diffusion dynamics considered in this paper not only displays the Rayleigh-Plateau instability similar to the Gregory-Laflamme instability but also has some intrinsic aspects each of which has a counterpart in the black hole dynamics. Of course, it could not be expected in general that black branes behave like the metal surfaces. However, the similarity of phase structures and the correspondences of dynamics will be worth further consideration. It is interesting to investigate these correspondences from a more fundamental perspective (cf.~Ref.~\cite{Kovtun:2003wp}).

As mentioned in Introduction, the behavior near the breakup and a coalescent process in the surface diffusion have been known to be described by self-similar solutions~\cite{BBW,Egg}. The self-similar solutions play important roles also in the gravitational collapse, especially in the critical phenomena~\cite{chop2,koike} and naked singularity formation~\cite{ori,harada}. We saw that depending on the parameter $p$ in the initial condition~(\ref{eq:IC}), the unstable uniform bridge will either pinch off or converge to the non-uniform bridge. This is analogous to the critical phenomena in the gravitational collapse~\cite{chop2}. It is interesting to investigate the present system from this point of view and to look for a counterpart in the black hole-black string system.

%%%-----------------------------------------------------------------%%%
%%%-----------------------------------------------------------------%%%
\medskip
\section*{Acknowledgments}
I would like to thank Oscar Dias, Roberto Emparan, Miyuki Koiso, Barak Kol, Kei-ichi Maeda, Mu-In Park, and Jennie Traschen for valuable comments and discussions. This work was supported by the Golda Meir Fellowship, Israel Science Foundation Grant No.607/05, and DIP Grant H.52.
%%%-----------------------------------------------------------------%%%
%%%-----------------------------------------------------------------%%%

\appendix

%----------------------------------------------------------------------%
%----------------------------------------------------------------------%
\section{ Calculation Details }
\label{sec:calc}
%----------------------------------------------------------------------%
%----------------------------------------------------------------------%

We derive the geometric quantities and some identities used in Sec.~\ref{sec:diff}.

%----------------------------------------------------------------------%
%----------------------------------------------------------------------%
\subsection{ Geometric Quantities }
\label{sec:calc}
%----------------------------------------------------------------------%
%----------------------------------------------------------------------%

A line element of $(n+2)$-dimensional flat space in cylindrical coordinates is given by
\begin{eqnarray}
	g_{\mu\nu} d x^\mu d x^\nu
	=
	d R^2 + d z^2 + R^2 d \Omega_n^2\ ,
\end{eqnarray}
where $\mu,\nu=1,2,\ldots,n+2$ and $ d \Omega_n^2 := \sigma_{ij}d y^i d y^j$, ($i,j=1,2,\ldots,n$) is the line element of a unit $n$-sphere. Let us consider an axisymmetric hypersurface $\Sigma$ given by $ \Phi (R,z) := R - r(z) = 0  $.
Since $d R=r^\prime d z$ on $\Sigma$, we have the induced metric $h_{ab}$, ($a,b=1,2,\ldots,n+1$) on $\Sigma$ and its determinant
\begin{eqnarray}
	h_{ab} d x^a d x^b
	=
	( 1+ r^{\prime 2} ) d z^2 + r^2 d \Omega_n^2\ ,
\;\;\;\;\;
	\sqrt{ h }
	=
	r^n \sqrt{ ( 1+ r^{\prime 2} ) \sigma }\ ,
\end{eqnarray}
where $\sigma:=\det \sigma_{ij}$. The Laplacian on $\Sigma$ is given by a formula
\begin{eqnarray}
	\Delta_{s}
	=
	\frac{ 1 }{ \sqrt{ h } }
	\partial_a \left( \sqrt{ h } \; h^{ab} \partial_b \right)\ .
\end{eqnarray}
If we assume that the functions to be differentiated depends only on $z$, we have $ \Delta_{s} $ in Eq.~(\ref{eq:nabla}).

An outward unit normal of $\Sigma$, denoting it by $n^\mu$, is given by
\begin{eqnarray}
	n_\mu
	=
	\frac{ \partial_\mu \Phi }{ \sqrt{ g^{\mu\nu} \partial_\mu \Phi \;  \partial_\nu \Phi  } }
	=
	\frac{ 1 }{ \sqrt{ 1+ r^{\prime 2} } }
	( \delta_\mu^R -r^\prime \delta_\mu^z )\ .
\end{eqnarray}
The mean curvature can be defined as the trace of an extrinsic curvature of $\Sigma$, which we denote by $K^\mu_\nu$. That is,
\begin{eqnarray}
	\kappa
	:=
	K^\mu_\mu
	=
	\nabla_\mu n^\mu
	=
	\left.
		\frac{ 1 }{ \sqrt{ g } }
		\partial_\mu \left( \sqrt{ g } \; g^{\mu\nu}
		n_\nu \right)
	\right|_{R=r(z)}
=
	\frac{ n }{ r \sqrt{ 1+ r^{\prime 2} } }
	-
	\partial_z
	\left( 
		\frac{ r^\prime }{ \sqrt{ 1+ r^{\prime 2} } } 
	\right)\ .
\ee

Finally, the normal velocity of $\Sigma$, $u$, is the projection of the velocity of the surface, which we denote by $v^\mu$, on the unit normal $n^\mu$, i.e., $ u := v^\mu n_\mu$.
Noting that $v^\mu = \partial_t r \delta^\mu_R $, we have the expression in Eq.~(\ref{eq:nabla}).

%----------------------------------------------------------------------%
%----------------------------------------------------------------------%
\subsection{ Area Decreasing }
\label{sec:time}
%----------------------------------------------------------------------%
%----------------------------------------------------------------------%

A few identities which might be useful to derive Eq.~(\ref{eq:dot}) are presented. Defining
\begin{eqnarray}
	j
	:=
	-\frac{ \kappa^\prime }{ \sqrt{ 1+r^{\prime 2} } } \ ,
\end{eqnarray}
equation~(\ref{eq:eom}) is rewritten as
\begin{eqnarray}
	r^n \partial_t r
	=
	-B \partial_z \left( r^n j \right)\ .
\label{eq:eom2}
\end{eqnarray}
One can easily show an identity,
\begin{eqnarray}
	\partial_z
	\left(
		\frac{ r^n r^\prime }{ \sqrt{ 1+r^{\prime 2} } }
	\right)
	=
	\frac{ n r^{n-1} r^{\prime 2} }{ \sqrt{ 1+r^{\prime 2} } }
	+
	\frac{ r^n r^{\prime\prime} }{ ( 1+r^{\prime 2} )^{3/2} }\ .
\label{eq:id}
\end{eqnarray}
Eliminating the second derivative, $r^{\prime\prime}$, from Eqs.~(\ref{eq:kappa}) and (\ref{eq:id}), we have 
\begin{eqnarray}
	n r^{ n-1 } \sqrt{ 1+r^{\prime 2} }
	=
	r^n \kappa
	+
	\partial_z 
	\left( 
		\frac{ r^n r^\prime }{ \sqrt{ 1+r^{\prime 2} } }
	\right) \ .
\label{eq:identity}
\end{eqnarray}
One can show Eq.~(\ref{eq:avdot}) using Eqs.~(\ref{eq:eom2}) and (\ref{eq:identity}) via integration by parts.

%----------------------------------------------------------------------%
%----------------------------------------------------------------------%
\section{ Second-Order Perturbation }
\label{sec:second}
%----------------------------------------------------------------------%
%----------------------------------------------------------------------%

One can extend the linear perturbation in Sec.~\ref{sec:RP} to arbitrarily higher-order ones. First, let us expand $r(t,z)$ around the UB,
\be
	r(t,z) = \sum_{m=0}^{\infty} \varepsilon^m r_m(t,z)\ ,
\label{eq:exp}
\ee
where $r_0$ and $r_1(t,z)$ are given in Sec.~\ref{sec:RP}. Hereafter, we set $r_0=B=1$ for simplicity.
As we are concerned only with the backreaction of the linear perturbation $r_1(t,z)$, we set $r_m(0,z)=0$, $(m\geq 2)$. Here, we consider the second-order perturbation. Substituting the expansion (\ref{eq:exp}) into the basic equation, we have the following at $O(\varepsilon^2)$,
\be
	\partial_t r_2 + n r_2^{ \prime\prime } + r_2^{ (4) }
	=
	n
	\left[
		-(n-2) r_1^{\prime 2}
		+ ( 2 r_1 - r_1^{\prime\prime} ) r_1^{\prime\prime}
		- 2 r_1^\prime r_1^{(3)}
	\right]\ .
\ee
After some calculations using the Fourier and Laplace transformations with respect to $z$ and $t$, respectively, we have
\be
	r_2 (t,z)
	=
	&&
	\sum_{i=1}^{N}
	A_i \left( 1 - e^{ 2 \omega(k_i) t } \right)
	+
	\sum_{i=1}^{N}
	B_i \left( e^{ 2\omega(k_i)t } - e^{ \omega(2k_i)t  } \right) \cos (2 k_i z)
\nonumber
\\
&&
	+
	\sum_{1\leq j < j \leq N}
	C_{ij}^+
	\left(
		e^{ [ \omega( k_i ) + \omega( k_j ) ] t }
		-
		e^{ \omega( k_i + k_j )t  }
	\right)
	\cos ( k_i + k_j ) z
\nonumber
\\
&&
	+
	\sum_{1\leq j < j \leq N}
	C_{ij}^-
	\left(
		e^{ [ \omega( k_i ) + \omega( k_j ) ] t }
		-
		e^{ \omega( k_i - k_j )t  }
	\right)
	\cos ( k_i - k_j ) z\ ,
\label{eq:second}
\ee
where
\be
&&
	A_i
	:=
	\frac{ n }{ 4 } a_i^2\ ,
\;\;\;\;
	B_i
	:=
	\frac{ n ( n-4-3k_i^2 ) k_i^2  }{ 2 [ 2 \omega( k_i ) - \omega( 2k_i ) ]} a_i^2 \ ,
\nonumber
\\
&&
	C_{ij}^{\pm}
	:=
	\pm
	\frac{
		n[
			( n - 2 \mp k_i k_j  ) k_i k_j
			\mp
			( 1 \pm k_i k_j ) ( k_i^2 + k_j^2 )
		]
	}
	{
		\omega(k_i)+\omega(k_j) - \omega( k_i \pm k_j )
	} a_i a_j
	\ .
\ee
From the above result, one recognizes an interesting non-linear effect. If we take $k_i > \sqrt{n}$, ($1\leq i \leq N$, note that we set $r_0=1$), the linear perturbation and almost all terms in the second-order perturbation (\ref{eq:second}) are suppressed exponentially or remain to be constant at most. However, the term containing factor $e^{ \omega( k_i - k_j )t  }$ in Eq.~(\ref{eq:second}) grows if there is a couple of sinusoidal waves satisfying $| k_i - k_j | < \sqrt{n} $, ($ i \neq j $) because $ \omega( k_i - k_j ) > 0$ for such a couple. This fact leads to the interesting behavior that high wavenumber modes begin to grow after an initial decay followed by an incubation time~\cite{CFM}.

%----------------------------------------------------------------------%
%----------------------------------------------------------------------%

%----------------------------------------------------------------------%
%----------------------------------------------------------------------%

\end{document}